\documentclass[fleqn,usenatbib,useAMS]{mnras}

\usepackage{graphicx}	
\usepackage{amsmath}	
\usepackage{amssymb}	
\usepackage{multicol}        
\usepackage{bm}		
\usepackage{pdflscape}	

\usepackage{hyperref, tabularx}

\usepackage[T1]{fontenc}
\usepackage{ae,aecompl}

\usepackage{txfonts}

\newcommand{\pc}{{\rm pc}}
\newcommand{\msun}{{\rm M}_\odot}

\newcommand{\lvsun}{{\rm L}_{V,\odot}}
\newcommand{\s}{{\rm s}}
\newcommand{\gyr}{{\rm Gyr}}
\newcommand{\magrm}{{\rm mag}}
\newcommand{\kms}{{\rm km/s}}

\newcommand{\alosmax}{\max(a_{\rm los})}
\newcommand{\alos}{a_{\rm los}}
\newcommand{\mbh}{M_{\rm BH}}

\newcommand{\sigmalos}{\sigma_{\rm los}}
\newcommand{\sigmapm}{\sigma_{\rm pm}}
\newcommand{\rh}{r_{\rm h}}
\newcommand{\sigmav}{\sigma_V}
\newcommand{\pdotorb}{\dot{P}_{\rm orb}}
\newcommand{\porb}{{P}_{\rm orb}}
\newcommand{\pulsara}{PSR\,B1820--30A}
\newcommand{\lmxb}{4U\,1820--30}

\newcommand{\limepy}{{\sc limepy}}
\newcommand{\python}{{\sc python}}
\newcommand{\emcee}{{\sc emcee}}

\title[Mass models of NGC\,6624 without an intermediate-mass black hole ]{Mass models of NGC\,6624 without an intermediate-mass black hole}

\author[Gieles et al.]{Mark Gieles$^1$\thanks{E-mail: m.gieles@surrey.ac.uk}, Eduardo Balbinot$^1$, Rashid I.S.M. Yaaqib$^1$,Vincent H\'{e}nault-Brunet$^{2}$, \newauthor Alice Zocchi$^{3,4}$,
Miklos Peuten$^1$, Peter G. Jonker$^{2,5}$
  \\
$^1$ Department of Physics, University of Surrey, Guildford, GU2 7XH, UK\\
$^2$ Department of Astrophysics/IMAPP, Radboud University, PO Box 9010, 6500 GL Nijmegen, The Netherlands\\
$^3$ Dipartimento di Fisica e Astronomia, Universit{\`a} degli Studi di Bologna, viale Berti Pichat 6/2, I40127, Bologna, Italy\\
$^4$ INAF - Osservatorio Astronomico di Bologna, Via Ranzani 1, I40127 Bologna, Italy\\
$^5$ SRON, Netherlands Institute for Space Research, Sorbonnelaan 2, NL-3584 CA Utrecht, The Netherlands
}

\date{Accepted 2017 October 11. Received 2017 11; in original form 2017 August 10}

\pubyear{2018}

\begin{document}
\maketitle

\begin{abstract}
An intermediate-mass black hole (IMBH) was recently reported to reside
in the centre of the Galactic globular cluster (GC) NGC\,6624, based
on timing observations of a millisecond pulsar (MSP) located near the
cluster centre in projection. We present dynamical models with
multiple mass components of NGC\,6624 -- without an IMBH -- which
successfully describe the surface brightness profile and proper motion
kinematics from the {\it Hubble Space Telescope} ({\it HST}) and the stellar-mass
function at different distances from the cluster centre.  The maximum
line-of-sight acceleration at the position of the MSP accommodates the
inferred acceleration of the MSP, as derived from its first period
derivative. With discrete realizations of the models we show that the
higher-order period derivatives -- which were previously used to
derive the IMBH mass -- are due to passing stars and stellar
remnants, as previously shown analytically in literature. We conclude that there is no need for an IMBH to explain the
timing observations of this MSP.
\end{abstract}




\section{Introduction}
\label{sec:introduction}
Finding an intermediate-mass black hole (IMBH), or providing evidence
against the existence for these sought-after objects, would be an
important step forward in our quest to understand the formation of
super-massive black holes in the centres of galaxies.  If we
extrapolate the relation between black hole masses and host galaxy
properties \citep{2000ApJ...539L...9F} below the mass range over which
it was established, then IMBHs could lurk in globular clusters
(GCs). For decades, the search for IMBHs in GCs has been a
cat-and-mouse game, in which claims for IMBH detections
(\citealt*{1976ApJ...208L..55N, 2008ApJ...676.1008N}; \citealt{
  2011A&A...533A..36L}) were soon after rebutted, either by improved
data \citep{2010ApJ...710.1032A, 2013ApJ...769..107L}, or by more
plausible, alternative interpretations of the data
\citep*{1977ApJ...218L.109I, 2017MNRAS.468.4429Z}.

Deep radio observations put upper limits to the mass of putative 
  accreting IMBHs of several $100\,\msun$ \citep{2012ApJ...750L..27S}
in nearby GCs. Because of the absence of gas in GCs, IMBH searches
mostly rely on stellar kinematics and dynamical modelling. The
  challenge with this approach is that the signal of an IMBH in the
kinematics of the visible stars is similar to that of a population
of stellar-mass black holes (\citealt*{2013A&A...558A.117L};
\citealt{2016MNRAS.462.2333P}; \citealt*{zocchi17b}) or
radially-biased velocity anisotropy \citep{2017MNRAS.468.4429Z}.

Individual stars with velocities above the local escape velocity have
been found in the core of some GCs \citep{1991ApJ...383..587M,
  2012A&A...543A..82L}, potentially pointing at the action of an
IMBH. However, also for these observations there exist alternative --
more plausible -- interpretations, such as slingshots after
interactions with a binary star \citep{1991AJ....101..562L}, or
energetically unbound stars that are trapped for several orbits in
  the Jacobi surface before they escape \citep*{2000MNRAS.318..753F,
  2017MNRAS.466.3937C, 2017MNRAS.468.1453D}. The periods of stars that
are bound to an IMBH are of the order of kyr, therefore excluding
the possibility of resolving full orbits of stars around it, as is
done in the Galactic Centre \citep[e.g.][]{2005ApJ...628..246E}.

A convincing signal of an IMBH would be a measure of the gravitational
acceleration in its vicinity, which can be obtained with timing
  observations. \citet{2014ApJ...795..116P} analysed the orbital
  period, $\porb$, of the low-mass X-ray binary (LMXB) \lmxb\ that
  sits at $1.3\arcsec$ from the centre in projection of the bulge GC
  NGC\,6624.  If there are no intrinsic binary processes changing the orbital period, then the period derivate, $\pdotorb$, is due to a
  gravitational acceleration along the line-of-sight ($\alos$), which
  contributes to $\pdotorb$ as $\pdotorb/P = \alos/c$
  \citep{1987MNRAS.225P..51B, 1993ASPC...50..141P}, where $c$ is the
  speed of light and we assumed that a positive $\alos$ implies an
  acceleration away from the observer.  \citet{2014ApJ...795..116P}
  find a large, negative $\pdotorb/P=-1.7\pm0.1\times10^{-15}\,{\rm
    s}^{-1}$.  The authors consider various possible explanations for
  the large $\pdotorb$ of \lmxb, including an IMBH with a mass
  $\mbh\simeq19\,000\,\msun$. They argue, however, that an IMBH is not
  a likely explanation, because \lmxb\ is part of a triple system, and
  the triple would not survive the tidal interaction with the
  IMBH. They also consider a population of centrally concentrated dark
  remnants as the source of the acceleration and conclude that this is
  a more likely explanation than an IMBH.  Whereas this is a plausible
  scenario, the decreasing $\porb$ of \lmxb\ is not exceptional when
  compared to other LMXBs which reside in the field and not in GCs
  [for a recent overview see table 4 in
  \citealt{2017ApJ...841...98P}]. About half of the LMXBs show an
  orbital period decrease of a similar or even larger magnitude than
  \lmxb. This could indicate that these systems are also accelerated
  due to the presence of a third body.  However, several of the
  alternative explanations possible for the observed orbital period
  changes, such as non-conservative mass transfer, mass-loss from the
  companion star, spin-orbit coupling discussed for instance in
  \citet{2017ApJ...841...98P}, are viable explanations for \lmxb\ as
  well.

Another way of inferring acceleration with timing observations
is with millisecond pulsars (MSPs). The precision with which the 
  spin period $P$ can be derived, allows for precise measurements of
its time derivative, $\dot{P}$, and higher-order derivatives, using
baselines of several years.  \citet{2017MNRAS.468.2114P} report the
finding of an IMBH in NGC\,6624, based on timing observations of
\pulsara. This pulsar sits at $0.4\arcsec$ from the centre of the
cluster and the authors use radio observations obtained over a
baseline of more than 25 years to derive $P^{(n)}$, up to $n=4$
  and even an upper limit for $P^{(5)}$. Under the assumption that
the MSP is bound to a point-mass, they infer the five orbital
elements of a Kepler orbit.  The mass they derive for the companion
  depends on what is assumed for the contribution to $\dot{P}$ due to
  intrinsic spin-down.  The fact that there is only a limit for
  $P^{(5)}$ also causes the mass of the companion to depend on the
  adopted eccentricity of the orbit of \pulsara. The MSP timing data
  are consistent with a low-mass companion ($\sim1\,\msun$) and a
  moderate eccentricity ($\sim0.35$), or a highly eccentric orbit
  ($\gtrsim0.9$) around a massive companion, which they consider to be
  an IMBH with $\mbh\gtrsim7\,500\,\msun$. The authors use $\mbh$
  inferred from $\pdotorb$ of \lmxb\ by \citet{2014ApJ...795..116P} to
  argue that a low-mass companion of \pulsara\ is not stable against
  tidal disruption. Combining the timing observations of \lmxb\ and
  \pulsara, \citet{2017MNRAS.471.1258P} conclude that NGC\,6624 has an
  IMBH with $\mbh \simeq 20\,000\,\msun$. 

The presence of such a massive IMBH has several consequences for the
distribution of the stars in the cluster, which are not observed in
  NGC\,6624. First, the GC should have a large core radius
(relative to the half-mass radius, $\rh$) \citep{2007PASJ...59L..11H}.
This core inflation is already important for black holes with masses
of the order of one per cent of the cluster mass
\citep{2004ApJ...613.1143B}. However, NGC\,6624 is a core-collapsed
cluster with an unresolved core radius
\citep[e.g.][]{1986ApJ...305L..61D}, making it an unlikely candidate
to host an IMBH.  Second, the presence of an IMBH quenches mass
segregation among the stars \citep{2008ApJ...686..303G}, while
NGC\,6624 displays clear signatures of mass segregation
\citep[e.g.][]{2016ApJ...832...48S}.

These dynamical arguments against the presence of an IMBH then beg for
an alternative interpretation of the timing observations of \pulsara,
which is what we address in this paper. By comparing dynamical
models to the surface brightness, stellar-mass function (MF) and kinematic
data of NGC\,6624, we constrain the mass distribution in NGC\,6624 to
determine whether this can accommodate the MSP timing observations. In
Section~\ref{sec:methods} we present the data and models and the
results are given in Section~\ref{sec:results}. Our conclusions and a
discussion are presented in Section~\ref{sec:conc_disc}.

\section{Data and models}
\label{sec:methods}

\subsection{Data}

\subsubsection{Surface brightness profile}
\label{ssec:sb}
There are several surface brightness profiles of NGC\,6624 available
in the literature \citep*[e.g.][]{1995AJ....109..218T,
  2006AJ....132..447N}. The (ground-based) Trager et al. profiles are
quite different from the (space-based) profiles of Noyola \& Gebhardt
in the inner $\sim10\arcsec$.  This is most likely because these
  studies used different positions for the cluster centre in deriving
  the surface brightness profile. Because NGC\,6624 is core collapsed,
  the core radius is not resolved and the surface brightness profile
  is sensitive to what is assumed for the
  centre. \citet{2010AJ....140.1830G} present updated positions for
  the centres of 65 Milky Way GCs using {\it Hubble Space Telescope} ({\it HST})
  data and an ellipse-fitting method. Based on the pulsar coordinates
  of \citet{2012ApJ...745..109L}, \pulsara\ is at a projected distance
  of $0.41\pm0.09\arcsec$ from this centre found by
  \citet{2010AJ....140.1830G}, where the uncertainty in the projected
  distance is due to the uncertainty in the cluster centre.  Because
  the inner surface brightness profile is important for constraining
  the mass profile near the MSP, we decided to re-derive the surface
  brightness profiles from archival {\it HST} WFPC2 data (prop-ID 5366)
  using the centre found by \citet{2010AJ....140.1830G}.  The images
consist of short (25s), medium (350s), and long (500s) exposures using
the F555W filter. For the purpose of deriving the surface brightness,
we use the short exposure only, as the brightest cluster stars are not
saturated.

To avoid geometric distortions in the WFPC2 data, we adopt the
\texttt{\_drz} images, where the field-of-view (FoV) has been
corrected for aberrations and the scale is homogeneous. The surface
brightness profile is built by summing the flux inside concentric
rings and dividing by the area, where the area is computed as the
total number of non-bad pixels within a given ring. Bad pixels are
defined as chip gaps, cosmic rays, or as being outside the image
boundary. We estimate the sky flux contribution in an uncrowded region
and subtract from the integrated flux.  The photometric uncertainties
are derived from the signal to noise ratio (SN), where we assume that
most of the noise comes from the sky contribution. Finally, we correct
for an extinction of $0.87\,{\rm mag}$ \citep{2010arXiv1012.3224H}.  
  Because the {\it HST} filter F555W is not exactly the same as the
  Johnson-Cousins $V$-band, we apply a scaling such that the surface
  brightness profiles matches those of Trager et al. and Noyola \&
  Gebhardt in the outer parts. As a check, we integrate the profile to
  obtain a total luminosity of $M_V=-7.49$, which matches the value
  quoted in \citet{2010arXiv1012.3224H}. From hereon we refer to our
  photometric system as $V$-band. The uncertainty in the magnitude is
derived assuming the signal dominated regime as $\sigmav= 2.5
\log_{10}(1 + 1/{\rm SN})$.

\begin{figure}
\includegraphics[width=8cm]{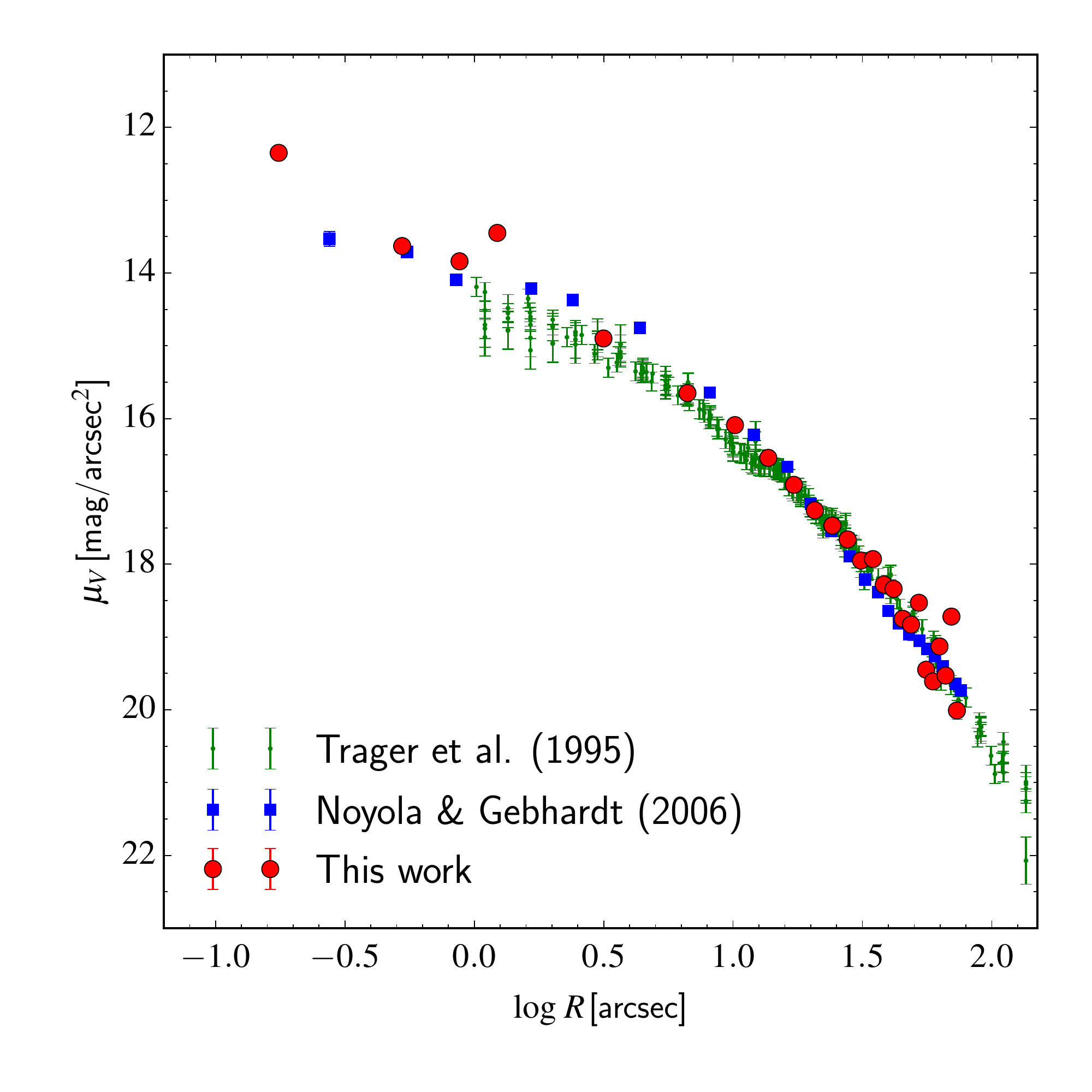}
\caption{Comparison between the surface brightness profiles available in literature and the one derived in this work.}
\label{fig:sb_comparison}
\end{figure}

In Fig.~\ref{fig:sb_comparison} we show the surface brightness profile derived in this work compared to the literature results of  Trager et al.  and Noyola \& Gebhardt. Because of the refined definition of the centre, we are able to get a value for $\mu_V$ at a closer distance to the cluster centre in projection ($R\simeq0.2\arcsec$) than Noyola \& Gebhardt, which is important to constrain the inner mass distribution of the cluster.

\subsubsection{Kinematics}
To constrain the mass of the GC, we use the one-dimensional velocity
dispersion ($\sigmapm$)  derived from {\it HST} proper motions
presented in \citet{2015ApJ...803...29W}. These data include stars
down to a magnitude 1 magnitude below the turn-off, and in the
modelling we assume that all stars for which we have velocities have
the same mass, equal to the turn-off mass.  We adopt a distance to
NGC\,6624 of 7.9\,kpc \citep{2010arXiv1012.3224H} to convert model
velocities to observed proper motion units of mas/yr.

There are only a few line-of-sight velocity measurements available for
stars in NGC\,6624 \citep{1989AJ.....98..596P,
  1992A&A...258..302Z}. The available data are not sufficient to
provide significant constraints on the line-of-sight velocity
dispersion profile ($\sigmalos(R)$). We therefore do not include these
data in the fitting, but compare our models to these data in Section~\ref{ssec:compare}. We note that \citet*{2011MNRAS.414.2690V}
present high-resolution spectroscopy of five stars in NGC\,6624, and
report line-of-sight velocities, but do not quote uncertainties and we
can therefore not use them.

\subsection{Models}
\subsubsection{Dynamical models}
We use the \limepy\ family of dynamical models
\citep{2015MNRAS.454..576G}\footnote{A \python\ implementation of the
  models is available from
  \href{https://github.com/mgieles/limepy}{https://github.com/mgieles/limepy}},
which are distribution function-based models that approximate
isothermal spheres in the centre and have a polytropic truncation near
the escape energy, making them suitable to describe dynamically
evolved and tidally limited systems, such as GCs.  The `sharpness' of
the truncation in energy is described by a parameter $g$, which
relates to the polytropic index $n$ as $n=g+1.5$. The concentration of
the model is determined by the dimensionless central potential $W_0$,
similarly to what is done in \citet{1966AJ.....71...64K} models (we
note that isotropic \limepy\ models with $g=1$ are indeed King
models.).  We adopt an isotropic velocity distribution, appropriate for
GCs in the late stages of their evolution (\citealt*{2016MNRAS.455.3693T}; \citealt{2016MNRAS.462..696Z}).  Multiple mass components can be included to
describe the effect of mass segregation \citep[as
  in][]{1979AJ.....84..752G}. The velocity scale of each mass
component ($s_j$) relates to the mass of the component ($m_j$)
as\footnote{We note that this relation results in equipartition at
  high masses, but the mass dependence of the central velocity
  dispersion is shallower at low masses [see
  \citealt{2015MNRAS.454..576G, 2016MNRAS.458.3644B,2017MNRAS.470.2736P}
  for details].} $s_j\propto m_j^{-1/2}$.  Extensive testing of
  \limepy\ against the results of direct $N$-body simulations was done
  by \citet{2016MNRAS.462..696Z} for single-mass models and
  \citet{2017MNRAS.470.2736P} for multimass models. Of particular
  importance for this study is that \limepy\ models accurately
  reproduce the degree of mass segregation in multimass systems.  The
meaning of the model parameter $W_0$ depends on the definition of the
mean mass, for which two options are available in \limepy. We use the
global mean mass of the entire model, rather than the central density
weighted mean mass, as is done in \citet{1979AJ.....84..752G}. We
refer to \citet{2017MNRAS.470.2736P} for a discussion on this choice.

\subsubsection{Stellar-mass function}
\label{ssec:mf}
\begin{figure}
\includegraphics[width=8cm]{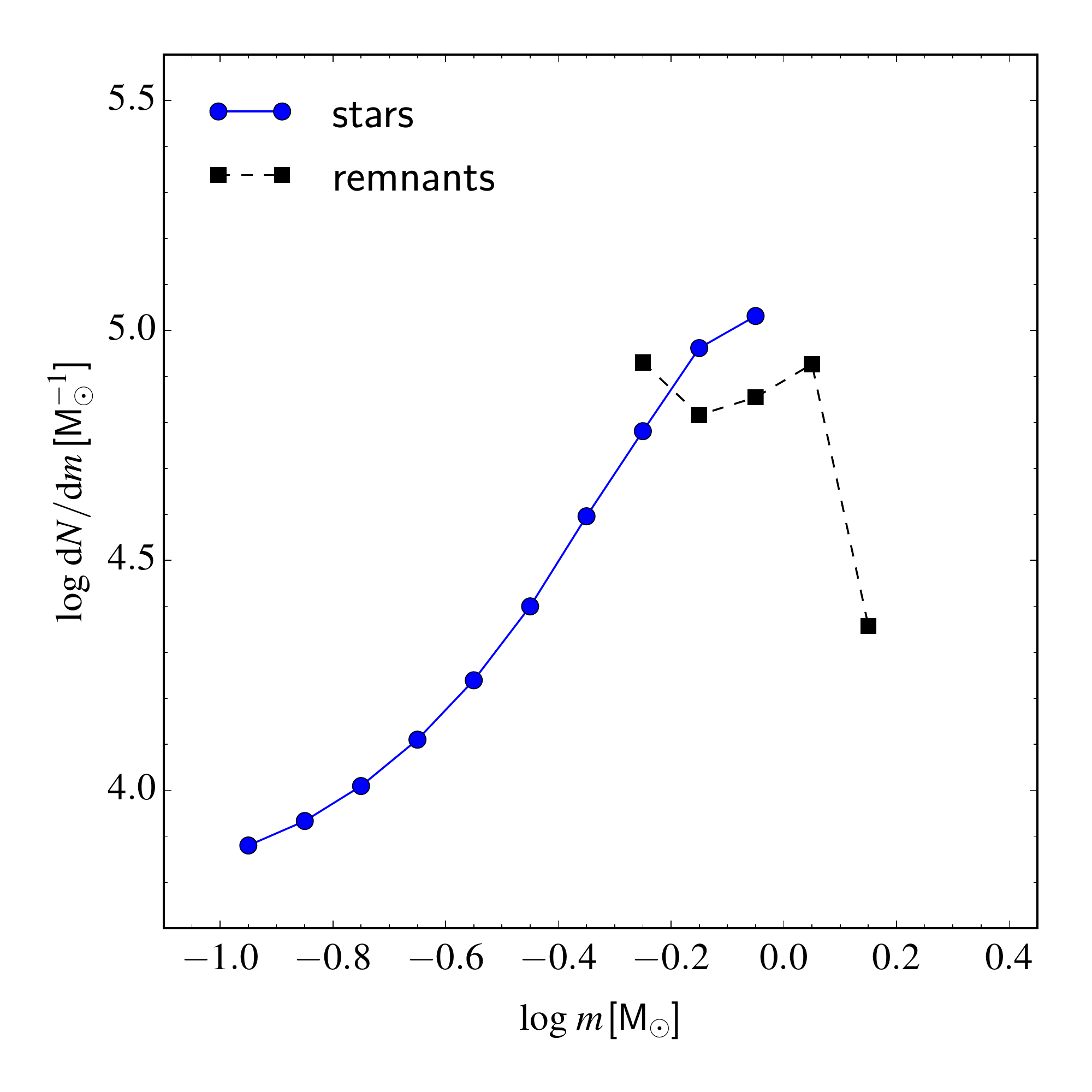}
\caption{The evolved stellar MF used to set up the multimass dynamical
  model. Because the cluster is near dissolution, the MF is depleted
  in low-mass stars and $\sim60$ per cent of the total mass resides in dark
  stellar remnants.}
\label{fig:mf}
\end{figure}

The multimass \limepy\ models require as input a mass function (MF)
for stars and remnants and for this we use the `evolved MF' algorithm
presented in \citet{balbinot17}. It assumes a
\citet{2001MNRAS.322..231K} stellar initial mass function (IMF), which
is then evolved for 12 Gyr by applying the effect of mass loss by
stellar evolution and remnant creation and the preferential escape of
low-mass stars and remnants as the result of evaporation. Based on the
orbit and the age of NGC\,6624, \citet{balbinot17} estimate that the
fraction of remaining cluster mass of NGC\,6624 is
$\mu=0.09\pm0.02$. We explored various MFs for different values of
$\mu$ and found good agreement with the observed stellar MF presented
by \citet{2016ApJ...832...48S} for $\mu = 0.065$, which we use from
hereon.  We adopt a retention fraction of neutron stars of 5 per cent and
assume that there are no stellar-mass black holes, as expected for
core collapsed clusters \citep{2013MNRAS.432.2779B}.  We note that
  the MF is relatively insensitive to the adopted neutron star
  retention fraction, because the total mass fraction in neutron stars
  for 100 per cent retention is only 2 per cent. For this MF, the fraction of the
total cluster mass in dark remnants is $0.62$. The MF of stars and
stellar remnants is shown in Fig.~\ref{fig:mf}. There are 10 mass bins
for the stars, and five mass bins for the remnants, such that there are
15 components in the multimass model.  For the stars, we convert mass
to $V$-band luminosity using the flexible stellar population synthesis
(FSPS) models of \citet{2010ApJ...712..833C}, adopting $Z= 0.0049$ and
an age of 12\,\gyr.

\begin{figure*}
\includegraphics[width=\textwidth]{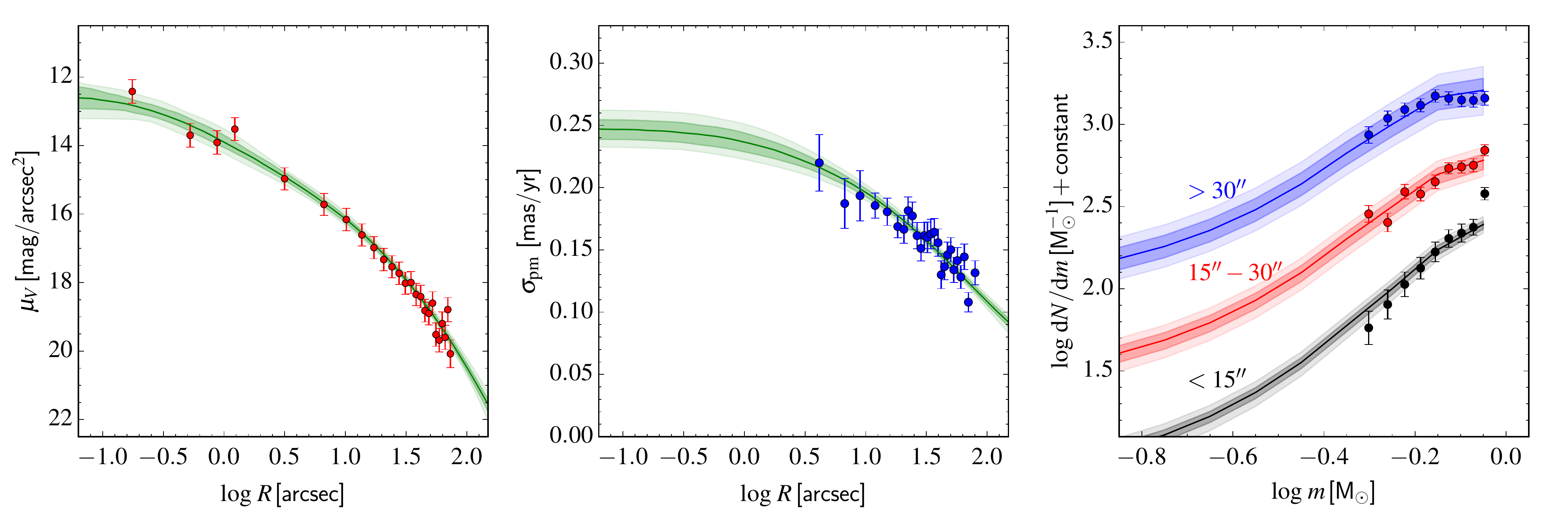}
\caption{Comparison between the observational data of NGC\,6624 and
  the dynamical models. The left panel shows the surface brightness
  profile, where the error bars are found from the (quadratic) sum of
  the uncertainty in the data and the nuisance parameter resulting
  from the fit: $\sigma_\mu \simeq 0.3\,\magrm/{\rm arcsec}^2$. The
  one dimensional velocity dispersion is shown in the middle panel and
  the right panel shows the stellar MF in different radial bins. The
  three MFs are shifted with arbitrary constants for clarity.}
\label{fig:panel}
\end{figure*}
\subsection{Model fitting}
With the MF defined, there are six fitting parameters: the
\limepy\ model parameters $W_0$ and $g$ and two  physical scales of
  the cluster, for which we use $\rh$ and the total cluster mass $M$.
Additionally, we fit on the global mass-to-light ratio $\Upsilon_V$,
which we use to scale the normalized luminosity profile to the surface
brightness profile. In this way, $M$ is only constrained by the
kinematics. Finally, we add a nuisance parameter $\sigma_\mu$ in
quadrature to $\sigmav$ to account for the effect of stochastic
sampling of the stellar luminosity function. We adopt uniform priors
for all parameters with the following range: $1\le W_0\le 12$, $0\le
g\le 2.5$, $10^4\le M/\msun\le 10^6$, $0.1\le \rh/\pc\le 20$, $0<
\Upsilon_V/(\msun/\lvsun)\le 5$ and $0<\sigma_\mu\le1$.  To determine
the posterior distributions of the model parameters and best-fit
values, we use the software package
\emcee\ \citep{2013PASP..125..306F}, which is a
pure-\python\ implementation of the Goodman \& Weare's affine
invariant Monte Carlo Markov Chain (MCMC) ensemble sampler
\citep{2010CAMCS...5...65G}. We use 200 walkers and after a few
hundred steps the fit converged. We continued for 2000 steps and in
the analyses we use the final walker positions to generate posterior
distributions and generate model properties. The
\python\ implementation of \emcee\ makes it straightforward to couple
it with \limepy.

\section{Results}
\label{sec:results}

\subsection{Comparison to data and model parameters}
\label{ssec:compare}
The resulting surface brightness and velocity dispersion profiles of
the models are compared to the data in the left and middle panels of
Fig.~\ref{fig:panel}, respectively. The median model values at each
radius are depicted with lines, and the $1\sigma$ and $2\sigma$
spreads are shown with dark and light shaded (green) regions,
respectively. The resulting stellar MFs at three distances from the GC
centre are shown together with the data of \citet{2016ApJ...832...48S}
in the right panel. Note that we did not fit on the MF. The spread in
the model MFs is the result of the variations in the model parameters,
leading to different MFs in the three regions.  The best-fit model
parameters and corresponding uncertainties (i.e. the median and
$1\sigma$ uncertainties) of the six parameters are given in
Table~\ref{tab:parameters}.

\setlength{\extrarowheight}{2pt}
\begin{table}
\caption{Summary of the fit results.}
\label{tab:parameters}
\begin{tabularx}{\columnwidth}{cccccc}
\hline
 $W_0$  & $g$ & $M$ & $r_{\rm h}$  & $\Upsilon$ & $\sigma_\mu$ \\ 
        &     & $[10^5\,\msun]$ & [pc]        & $[\msun/\lvsun]$ \\ \hline
        \vspace{0.1cm}
  $ 9.83_{-1.30}^{+1.10}$ &  $ 2.24_{-0.12}^{+0.18}$ &  $ 1.11_{-0.13}^{+0.12}$ &  $ 2.40_{-0.58}^{+0.42}$ &  $ 1.30_{-0.09}^{+0.09}$ &  $ 0.32_{-0.07}^{+0.05}$ \vspace{0.05cm}\\
\hline
\end{tabularx}
\end{table}

\begin{figure}
\includegraphics[width=8cm]{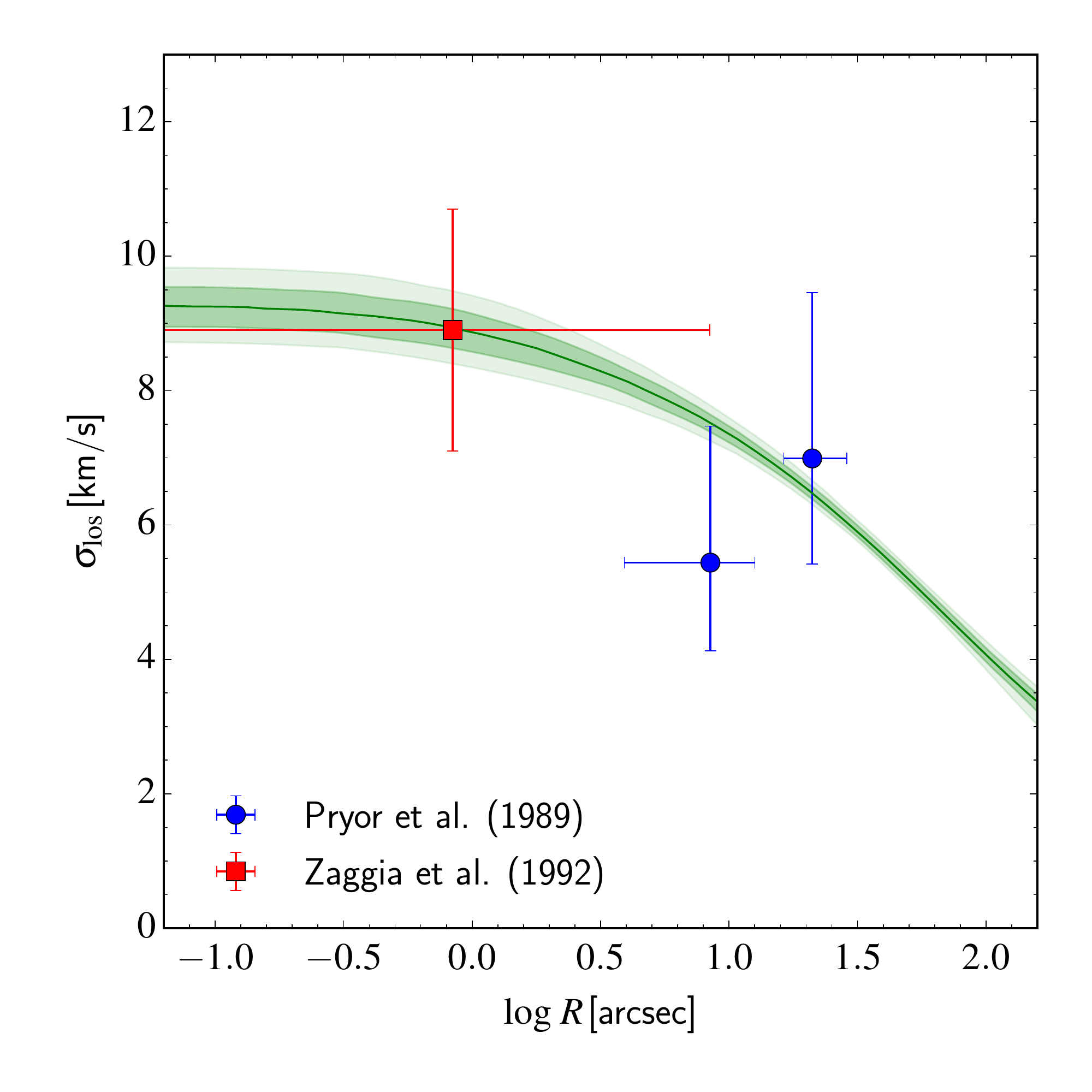}
\caption{Line-of-sight velocity dispersion from the dynamical models
    compared to literature data.}
\label{fig:vrms}
\end{figure}

\citet{1992A&A...258..302Z} measured $\sigmalos$ in the centre of
NGC\,6624, using integrated light spectroscopy of the inner
$8.4\arcsec\times4.6\arcsec$ and find
$\sigmalos=8.9\pm1.8\,\kms$. \citet{1989AJ.....98..596P} present
velocities of 19 stars, of which 18 stars between $\sim4\arcsec$ and
$\sim15\arcsec$ from the centre. We split these data in two samples of
9 stars (excluding their innermost isolated data point at
$R\simeq0.6\arcsec$), with respect to the median distance to the
centre and determine $\sigmalos$ from their line-of-sight velocities
via a maximum likelihood method using \emcee. The resulting
dispersions from the Pryor et al. data at two locations and the Zaggia
et al. measurement are compared to the velocity dispersion of our
models in Fig.~\ref{fig:vrms}. The large uncertainties in the measured
$\sigmalos$ (relative to the proper motion dispersion, see
Fig.~\ref{fig:panel}) do not allow us to use $\sigmalos$ to further
constrain the model.

\subsection{Line-of-sight acceleration}
\label{ssec:alos}
\begin{figure}
\includegraphics[width=8cm]{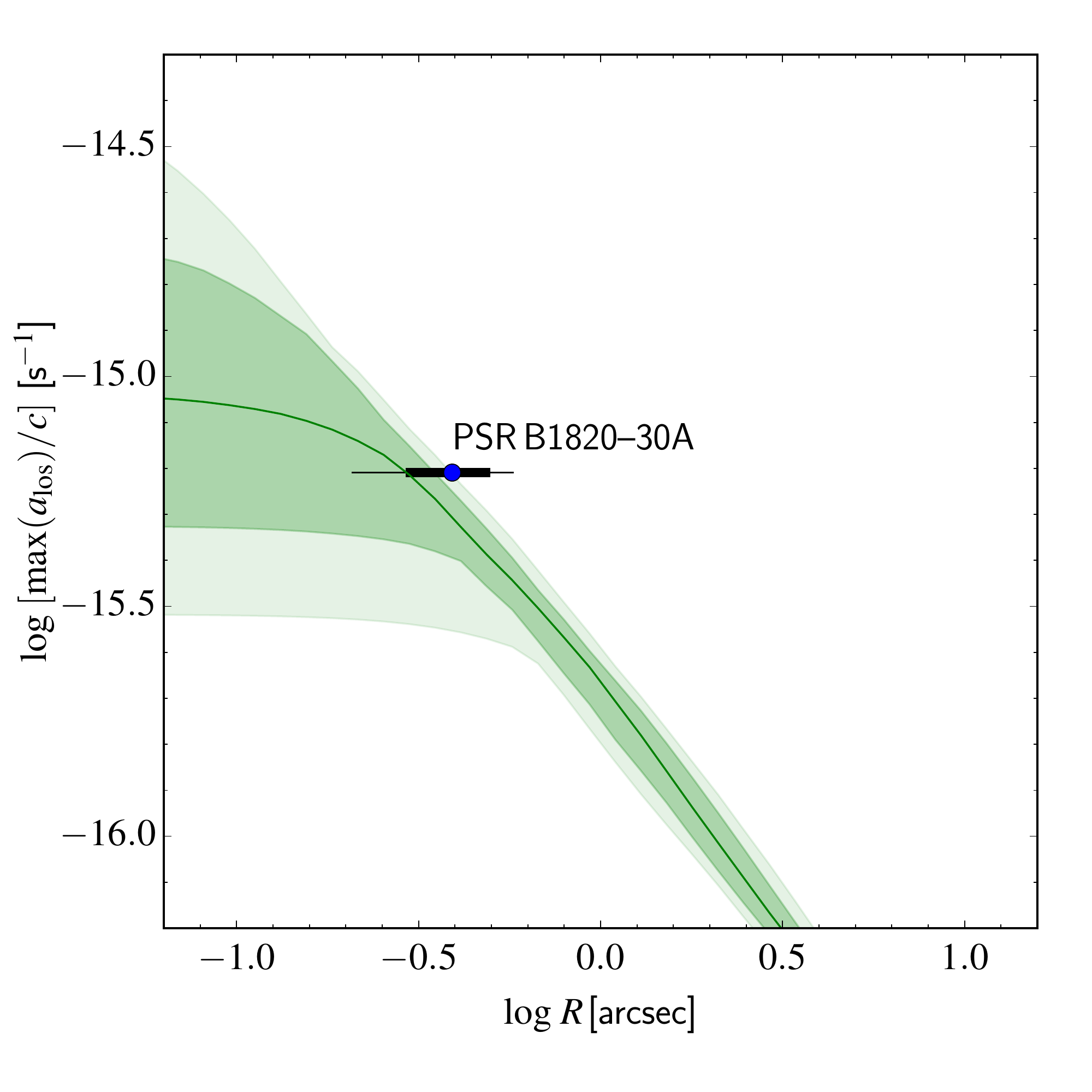}
\caption{Comparison of the maximum acceleration along the
  line-of-sight as a function of distance to the centre in projection
  for the models and the inferred acceleration of \pulsara.  The
  median result of the models is shown as a line and $1\sigma$ and
  $2\sigma$ spreads are shown as dark and light shaded regions. Within
  the $1\sigma$ uncertainty, the inferred acceleration of
  \pulsara\ can be accommodated by the enclosed mass profile of the
  cluster.}
\label{fig:acc}
\end{figure}

For each model we determine the maximum acceleration along the
line-of-sight as a function of distance to the centre in projection,
$\alosmax$. The result is shown in Fig.~\ref{fig:acc}, together with
the acceleration of \pulsara\ as inferred from $\dot{P}/P$. The
thick(thin) horizontal error bar shows the $1\sigma(2\sigma)$
uncertainties in the distance from the centre, due to the uncertain
position of the cluster centre (see Section~\ref{ssec:sb}). Accounting
for the uncertainties in the models and the pulsar position,
$\alosmax$ of the models can accommodate for the acceleration derived
from $\dot{P}/P$. \citet{1993ASPC...50..141P} showed that the maximum
acceleration in the core is proportional to the central surface mass
density ($\Sigma_0$), with $\alosmax/c \simeq5\times10^{-16}\,\s^{-1}$
for $\Sigma_0=10^6\,\msun/\pc^2$. From the models we derive $\Sigma_0
= 1.86_{-0.90}^{+2.22}\times10^6\,\msun\,\pc^{-2}$, such that the
observed $\dot{P}/P \simeq 6.2\times10^{-16}\,\s^{-1}$ is comfortably
below the expected maximum inside the core [$\alosmax/c
\simeq10^{-15}\,\s^{-1}$].

The central mass-to-light ratio in the $V$-band is $\Upsilon_{V,0} =
5.47_{-1.60}^{+3.33}\,\msun/\lvsun$, larger than the global
$\Upsilon_V\simeq1.3\,\msun/\lvsun$ (see
Table~\ref{tab:parameters}). This is because of the central
concentration of dark remnants. From this we see that the pulsar
acceleration can be explained by mass models, based on a canonical IMF
evolved to an age of 12 Gyr for the effects of stellar evolution and
the escape of low-mass stars (see Section~\ref{ssec:mf} and
Fig.~\ref{fig:mf} for details), without the need for an IMBH.

\subsection{Discrete models}
We quantify the effect of nearby passing stars and remnants on the
spin derivatives by generating $10^3$ discrete realizations of the
  multimass models from randomly drawn walkers of the final MCMC
chains. In each model we add $10^3$ mass-less tracers, at positions
corresponding to \pulsara. We assign projected distances on a ring
with radius $0.41\pm0.09\arcsec$, assuming a Gaussian spread, and we
sample positions along the line-of-sight from the density
distribution of the neutron stars. We then use the expressions for the
acceleration, jerk, snap and crackle from \cite{2008NewA...13..498N},
to derive $P^{(n)}$, with $1\le n\le4$, respectively. We omit
$P^{(5)}$ because there is only an upper limit available.

\begin{figure*}
\includegraphics[width=\textwidth]{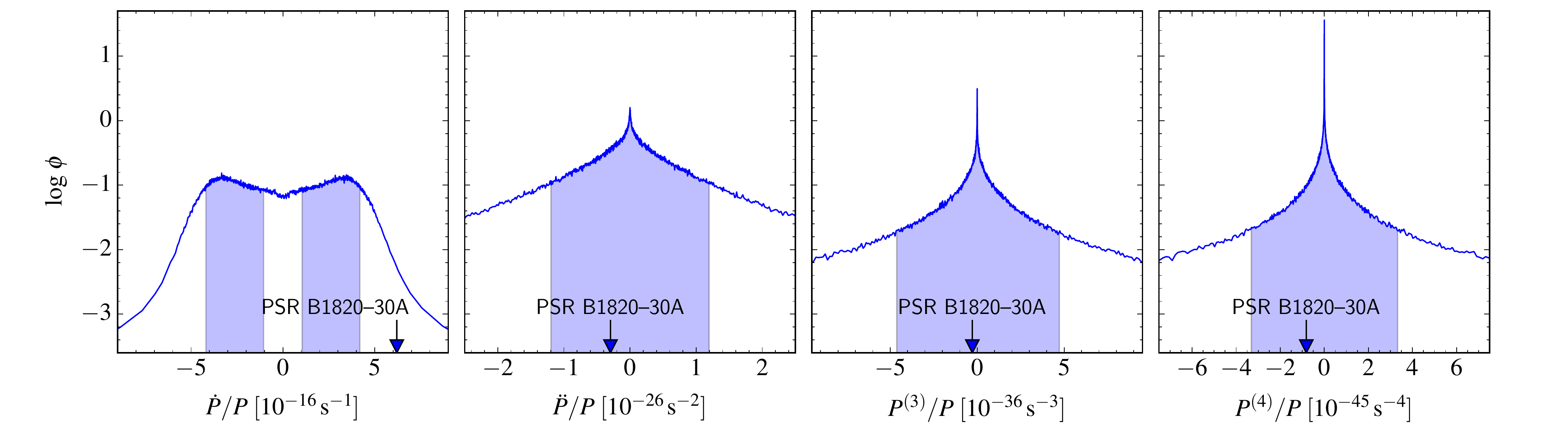}
\caption{Frequency $\phi$ of $P$ derivatives from discrete
  realizations of the dynamical models. Observed values from timing
  observations of \pulsara\ by \citet{2017MNRAS.468.2114P} are
  indicated with arrows. The shaded regions indicate the values
    between the 16 and 84 percentiles, around the most likely
    value(s).}
\label{fig:pderivs}
\end{figure*}

In Fig.~\ref{fig:pderivs} we show the frequency ($\phi$) for each of
the $P^{(n)}$ for the $10^6$ tracers. The regions containing $68\%$ of
the points (i.e. between the 16 and 84 percentiles) are
indicated with a (blue) shaded area.  For $\dot{P}$, the distribution
peaks near the maximum value, which means that it is more likely to
find an acceleration near the maximum, than near 0.

The observed values for \pulsara\ (derived from the frequencies
reported by \citealt{2017MNRAS.468.2114P}) are indicated with an arrow
in each panel. The $\dot{P}$ of \pulsara\ is slightly beyond the peak
of the distribution, with $1.5\%$ of the tracer particles having
$|\dot{P}|/P$ larger than observed. \citet{1987MNRAS.225P..51B} showed
that the typical contribution of passing stars to $\ddot{P}$ is of
comparable magnitude as the contribution of the mean gravitational
field and \citet{1993ASPC...50..141P} showed that this is also true
for the higher-order derivatives. We confirm this here: the
higher-order derivatives of \pulsara\ are within the $1\sigma$
spreads, implying that values of $P^{(n)}$ with $n\ge2$ are dominated
by stochastic effects and can not be used to infer the smooth
underlying potential.

By considering the correlations between $P^{(n)}$ with $n\ge2$ and
$\dot{P}$, we find that for the points with the largest $|\dot{P}|/P$,
the higher-order derivatives are also near the maximum of their
respective distributions.  This is because the derivatives for these
points are dominated by a single, nearby star rather than the global
potential. For a value of $|\dot{P}|/P \simeq 6.2\times10^{-16}\,{\rm
  s}^{-1}$, as found for \pulsara, the most likely values of
$|P^{(n)}|/P$ for $n=[2,3,4]$ are roughly $[5\times10^{-26}\,{\rm
    s}^{-2}, 5\times10^{-35}\,{\rm s}^{-3}, 10^{-44}\,{\rm s}^{-4}]$,
all well outside the $1\sigma$ regions indicated in
Fig.~\ref{fig:panel}.  The observed values of $P^{(n)}$ with $n\ge2$
are within the $1\sigma$ intervals of our models, which suggests that
$\dot{P}$ is not the result of the acceleration due to a single,
nearby star and an additional intrinsic spin-down contribution to
$\dot{P}$ is required.  For $|\dot{P}|/P\simeq4\times10^{-16}\,{\rm
  s}^{-1}$, the higher-order derivatives can have any value shown in
the distributions in Fig.~\ref{fig:pderivs}. This suggests that at
least $\sim30$ per cent of $\dot{P}$ is due to intrinsic spin-down.  We
discuss in Section~\ref{sec:conc_disc} that such a contribution to
$\dot{P}$ is not unreasonable given the high $\gamma$-ray luminosity
of \pulsara. Such a contribution to $\dot{P}$ would also make the
resulting line-of-sight acceleration easier to explain by our models,
because $\sim20$ per cent of our model points in the left panel of
Fig.~\ref{fig:pderivs} have $|\alos|/c\gtrsim4\times10^{-16}\,{\rm
  s}^{-1}$.

\vspace{0.5cm}
\section{Conclusions and discussion}
\label{sec:conc_disc}
\begin{figure}
\includegraphics[width=8cm]{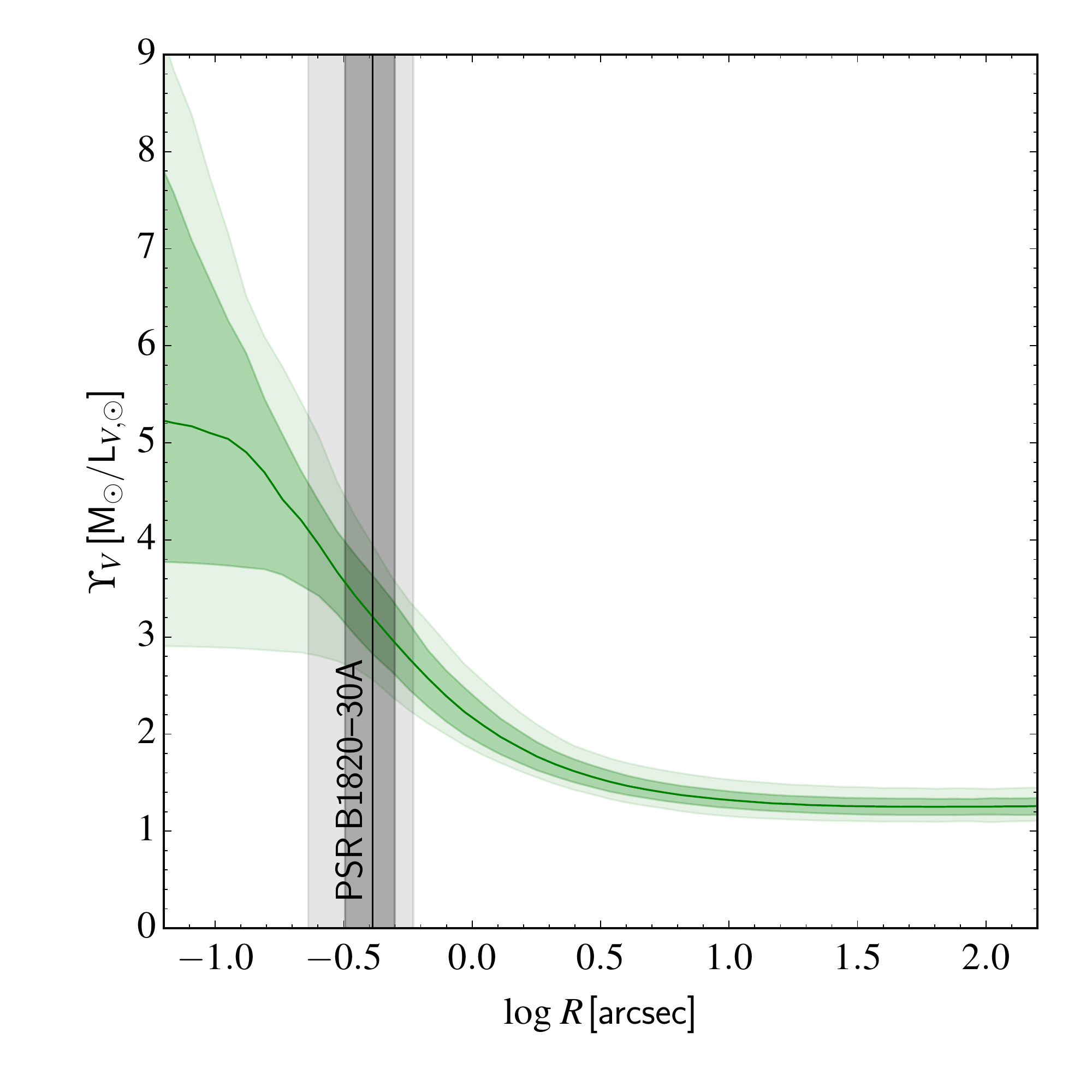}
\caption{{ Mass-to-light ratio profile from the dynamical models. The
    location of \pulsara\, and corresponding $1\sigma$ and $2\sigma$
    uncertainties are indicated with a vertical line and (black)
    shaded regions, respectively. }}
\label{fig:ml}
\end{figure}

\citet{2017MNRAS.468.2114P,2017MNRAS.471.1258P} conclude that an IMBH
with $\mbh\gtrsim7\,500\,\msun$ is required to explain the timing
observations of the MSP in the core of NGC\,6624.  We have shown that
$\dot{P}$ and higher-order derivatives of $P$ of \pulsara\ in
NGC\,6624 can be explained by dynamical multimass models without an
IMBH.  The models were derived from fits to the surface brightness and
kinematics profiles of this GC.  The best-fit dynamical models have
central densities of $\rho_0 =
7.54_{-5.56}^{+34.3}\times10^7\,\msun\,\pc^{-3}$ and a central surface
density of $\Sigma_0 =
1.96_{-0.98}^{+2.51}\times10^6\,\msun\,\pc^{-2}$, which explains the
high acceleration of the MSP near the centre. Although these central
densities are high, they are in the range expected for core-collapsed clusters. For
example, \citet{2014MNRAS.438..487D} find
$\rho_0\simeq3\times10^7\,\msun\,\pc^{-3}$ in Jeans models of
M15. Similar values have also been found in evolutionary models of
core-collapsed clusters: \citet{1992ApJ...392...86G} modelled
NGC\,6624 with Fokker--Planck models and find
$\rho_0\simeq2\times10^7\,\msun\,\pc^{-3}$, which combined with their
core radius of $r_0\simeq0.05\,\pc$ gives rise to
$\Sigma_0=\rho_0r_0\simeq10^6\,\msun\,\pc^{-3}$. Similar results were
obtained with Fokker--Planck models of M15 \citep*{2011ApJ...732...67M}
and $N$-body models of NGC\,6397 \citep{2009MNRAS.397L..46H}. We
  note that the enclosed mass within the radius of the MSP is lower
  than the inferred IMBH mass of \citet{2017MNRAS.468.2114P}:
  $M(<0.41\arcsec) = 646_{-163}^{+74}\,\msun$ (and in three
  dimensions: $M(<0.41\arcsec)= 362_{-199}^{+93}\,\msun$). We
  therefore agree with the conclusion of \citet{2014ApJ...795..116P}
  that NGC\,6624 has a population of centrally concentrated dark
  remnants. However, these authors did not present dynamical
  models. They varied the inner (mass) density profile with respect to
  the light profile to obtain a central mass profile that could fully
  explain $\pdotorb$ of \lmxb\ by the line-of-sight
  acceleration. Their $\Upsilon_V(R)$ starts to significantly increase
  within $R\lesssim5\arcsec$, whereas in our models $\Upsilon_V(R)$
  rises only within $R\lesssim1\arcsec$ (see Fig.~\ref{fig:ml}). The
  line-of-sight acceleration of our models at the position of
  \lmxb\ ($R\simeq1.3\arcsec$) is therefore not able to explain
  $\pdotorb$ of \lmxb. New analyses show that the LMXB may be closer to the centre of the cluster (Jay Strader, private communication; Tremou et al., in preparation), but even if it resides as close to the centre of NGC\,6624 as \pulsara, only $\sim30$ per cent of $\pdotorb$ would be due to a gravitational acceleration. 
  More importantly, as we argued in
  Section~\ref{sec:introduction}, there are LMXBs in the field with
  similar (or higher) $|\pdotorb|$, which is why we do not interpret
  this signal as being the result of an acceleration. 
 
Our global mass-to-light ratio $\Upsilon_V =1.30\pm0.09\,\msun/\lvsun$
is lower than what is expected from stellar population models. From
the FSPS models \citep{2010ApJ...712..833C} we estimate
$\Upsilon_V\simeq2.88 \,\msun/\lvsun$, for a stellar population with
an age of 12\,\gyr, $Z=0.0049$ and a \citet{2001MNRAS.322..231K}
IMF. The difference is because NGC\,6624 is depleted in low-mass stars
(see the right panel in Fig.~\ref{fig:panel}) as the result of dynamical
evolution in the Galactic tidal field. By matching the present-day MF
of the FSPS model to the observed MF, we obtain
$\Upsilon_V=1.15\,\msun/\lvsun$, close to what we infer from the mass
models. This agreement lends further support to the validity of the
best-fit multimass models.

Our results depend on the distance $D$, for which we adopted
$D=7.9$\,kpc \citep{2010arXiv1012.3224H}. The model properties depend
on $D$ in the following way: the velocity dispersion in [km/s] derived
from the proper motions and $\rh$ in [pc] both depend linearly on
$D$. From virial equilibrium arguments, the total dynamical mass is
proportional to $\sigma^2\rh\propto D^3$.  The surface density
\citep[and therefore the line-of-sight acceleration; see
  Section~\ref{ssec:alos} and][]{1993ASPC...50..141P}, scales as
$D$. The inferred $\alosmax$ is therefore also proportional to $D$.
With line-of-sight velocities beyond $\gtrsim3\arcsec$ a dynamical
distance to NGC\,6624 can be obtained.  Improved line-of-sight
velocities of stars within $\lesssim3\arcsec$, i.e. where there are no
proper motions available, would also be helpful to place tighter
constraints on the mass profiles in the centre
\citep{1986AJ.....92...72R}.

From randomly selecting walkers from the MCMC chain and populating the
density profile of the neutrons stars with mass-less tracers, we
find that in about 1.5 per cent of cases the line-of-sight acceleration is
higher than what is inferred for \pulsara. Although this is suggestive
that the model (surface) densities are on the low-side to account for
the pulsar $\dot{P}$, we note that there is a contribution to the
observed $\dot{P}$ from intrinsic spin-down due to magnetic breaking
that we have so far not discussed.  The $\gamma$-ray data of
\cite{2011Sci...334.1107F} show that \pulsara\ dominates the
$\gamma$-ray emission from NGC\,6624. In fact, \pulsara\ is the most
luminous $\gamma$-ray MSP known. \cite{2011Sci...334.1107F} estimate
from the $\gamma$-ray emission that at least $10$ per cent of the observed
$\dot{P}$ is due to the intrinsic spin-down, when one assumes an
unrealistic $\gamma$-ray efficiency of $\eta = 1$. If one assumes a
more realistic efficiency of $\eta\simeq0.1$
  \citep{2011Sci...334.1107F}, the majority of the observed $\dot{P}$
could in fact be due to intrinsic effects.

There are five more pulsars known in NGC\,6624
\citep{2012ApJ...745..109L}, and for two of these there are $\dot{P}$
measurements. Both have $P\simeq0.4\,\s$, and their $\dot{P}$ values
are comparable to what is found for pulsars in the field with similar
$P$ \citep{2005AJ....129.1993M}, implying that these pulsars can not
be used to infer the gravitational potential.

MSPs are potentially the only way of inferring the acceleration with
sufficient precision to make a viable case for an IMBH. To form a MSP,
a pulsar needs to tidally capture a star, which subsequently spins up
the pulsar by angular momentum transport via Roche overflow
\citep{1987Natur.329..312V}. However, high stellar densities are
required for tidal capture to be efficient
\citep{1975MNRAS.172P..15F}, and an IMBH reduces the stellar density
\citep{2007PASJ...59L..11H}, making GCs with MSPs unlikely GCs to
possess an IMBH. Given the degeneracies of a dynamical signal of an
IMBH with radial velocity anisotropy \citep{2017MNRAS.468.4429Z} and
the presence of a stellar-mass black hole population
\citep{2013A&A...558A.117L,2016MNRAS.462.2333P,zocchi17b}, perhaps a
convincing detection will need to come from gravitational microlensing
\citep{2016MNRAS.460.2025K}.  Future instrumentation, such as the ELT
first-light instrument MICADO \citep{2010SPIE.7735E..2AD}, may be able
to resolve the proper motion velocity dispersion to sufficiently close
distances to the centre of clusters and for enough stars to be able to find convincing
signatures of IMBHs in the cores of GCs, if they exist.

\section*{Acknowledgements}
We thank Ben Perera for comments on an earlier version of this paper
  and Markus Kissler-Patig for a constructive referee report and
  suggestions that have helped us to improve the paper. We thank Denis
  Erkal for suggesting to look into correlations between the various
  $P^{(n)}$ in the discrete models. We thank members of the Surrey
  astrophysics group and the Gaia Challenge collisional working group
  for various discussions on mass modelling of gravitational systems.
MG acknowledges financial support from the Royal Society (University
Research Fellowship).  MG, EB and MP acknowledge support from the
European Research Council (ERC StG-335936, CLUSTERS).  VHB
acknowledges support from the Radboud Excellence Initiative
Fellowship. PGJ acknowledges support from the European Research
Council (ERC CoG-647208).



\bsp	
\label{lastpage}
\end{document}